\documentstyle[prl,aps,epsf,multicol]{revtex}
\begin{document}
\draft
%\twocolumn[\hsize\textwidth\columnwidth\hsize\csname@twocolumnfalse%
%\endcsname
 
\title{Critical Dynamics of Thermal Conductivity at the Normal-Superconducting Transition}

\author{Smitha Vishveshwara$^1$ and Matthew P. A. Fisher$^2$}
\address{$^1$Department of  Physics, University of California,
Santa Barbara, CA 93106 \\ 
$^2$Institute for Theoretical Physics, 
University of California, Santa Barbara, CA 93106--4030 
}

\date{\today}
\maketitle

\begin{abstract}
We study the effect of thermal fluctuations on the critical dynamics
of the three-dimensional disordered superconductor across the
normal-superconducting transition. We employ a phenomenological
hydrodynamic approach, and in particular find the thermal conductivity
 to be smooth and non-singular at the transition.
\end{abstract}

\vspace{0.15cm}

%\pacs{PACS numbers:75.10.Nr, 05.50.+q, 75.10.Jm}
%\vskip -0.5 truein 

\begin{multicols}{2}
\narrowtext

\section{Introduction} 

Some fifteen years after the discovery of the cuprate
superconductors, many tantalizing puzzles remain,
particularly in regard to the ground state properties
and possible quantum criticality as the doping is varied.
It has been
difficult to disentangle consistent analyses faithful to experiment
from the zoo of microscopic theories and models. 
In contrast, there have been notable successes
in understanding the phenomenology of the
finite temperature transition into the superconducting phase,
where quantum fluctuations are unimportant and progress
can be made employing classical Landau-Ginzburg approaches.
In particular, at optimal doping in YBCO,
the observed critical behavior 
in the electrical conductivity, specific heat,
penetration depth, and other quantities\cite{critex} 
 appears to be consistent with theoretical expectations.

Recent attention has focused on thermal transport experiments\cite{thermex},
which have shed light on the low temperature transport of quasiparticles
in the superconducting phase, and have revealed
low temperature violations
of the Wiedemann-Franz law in the normal state
of the electron doped material suggestive
of a non-Fermi liquid ground state\cite{Taill}.
Here, we revisit the theory of thermal conductivity,
focusing on the critical behavior near the finite
temperature superconducting transition where
progress is possible without the need for a microscopic
quantum model.  Older works within a BCS framework
generalizing the Aslamazov-Larkin calculations to
thermal conductivity\cite{AL}, have predicted a diverging thermal
conductivity upon cooling into the superconductor, reminiscent
of the behavior of He-4 at the $\lambda$-transition.
This appears to be at odds with experiment in the cuprates, which typically
show
a finite and rather smooth thermal conductivity as one cools
through $T_c$, with a large growth upon further cooling usually
ascribed to quasi-ballistic quasiparticle transport.

Our study focuses on the three-dimensional disordered superconductor,
most appropriate to optimally doped YBCO which is the least two-dimensional
cuprate.  We follow the 
phenomenological hydrodynamic approach to critical dynamics
pioneered by Halperin and Hohenburg\cite{HH}.  Indeed, one of the early
successes of this dynamical scaling approach was the correct
description
of the diverging thermal conductivity near the $\lambda$-transition
in He-4.  Here, we modify this theory to account for
impurities and long-ranged Coulomb forces appropriate
to the superconductor.  Our central conclusion is
that rather than a divergent thermal conductivity
as in He-4, the thermal conductivity
is predicted to be {\it finite} and exhibit {\it smooth} non-singular
behavior
at the superconducting transition.
Surprisingly, the critical singularities which dominate the electrical
conductivity and other thermodynamic properties are
found to be {\it completely} absent from the thermal conductivity.
There is effectively a decoupling between the
thermal and electrical transport coefficients upon approaching
the superconducting phase, in strong contrast to
the universal Lorenz ratio relating these two transport coefficients
in a conventional metal.

\section{Thermal Conductivity} 

The thermal conductivity $\kappa$, relates the 
heat current $\vec{Q}$ to an applied
thermal gradient $\vec{\nabla}T$ under the condition of no particle flow, 
\begin{equation}
\vec{Q} = - \kappa \vec{\nabla} T, \quad \vec{j} = 0,
\label{therm}
\end{equation}
where $\vec{j}$ is the particle current. 
Within linear response one has an
Einstein relation for the thermal conductivity,
\begin{equation}
\kappa = D_T C_V,
\label{Einstein}
\end{equation}
where $D_T$ is the thermal diffusion constant, and $C_V$ the specific heat
at constant volume.

In systems with a condensed ground state such as superfluid He-4, the situation is a little more
involved since energy can couple to the order parameter in a 
non-dissipative fashion.  This leads to a ballistic wave-like propogation
of heat  - second sound -  in the superfluid state
,
 and an infinite thermal conductivity.
In the normal state, while heat does propogate diffusively,
its associated thermal conductivity diverges due to critical
fluctuations upon approaching the transition temperature.  
As we shall discuss, in contrast to its sister system of
superfluid He-4, the impure superconductor can only support a 
diffusive rather than an oscillatory heat mode, and the thermal conductivity
$\kappa$ is finite at all temperatures. 

In the high $T_c$ 
superconductors, the role of thermal fluctuations near
the superconducting transition is drastically enhanced relative
to their low $T_c$ counterparts.  Being generally anisotropic,
and strongly type II,
the Ginzburg criterion shows that critical fluctuations
 are present over a relatively
wide range of temperatures, perhaps as large as 5-10 Kelvin\cite{FFH}.
Within this temperature window we can dispense with microscopic
models, and appeal to a critical hydrodynamic approach. 

\section{Superfluid hydrodynamics revisited} 

To identify the appropriate model near criticality,
we follow Ref.\cite{HH}, and first
focus on the hydrodynamics in the normal state.
Since the frequency $\omega$ of the 'slow' hydrodynamic modes
vanishes 
with wave-vector $\vec{q}$, one need only study the dynamics of
 conserved densities which cannot relax rapidly on long lengthscales.
We consider first an uncharged and pure fluid, such
as He-4.
In this case
there are five such conserved densities:
three components of the particle current density
 ($\vec{j}$) and the energy ($\epsilon$), in addition
to the particle number   density ($\rho$).  Associated with these
five conserved densities are five hydrodynamic modes:
propagating
first sound which involves the density and longitudinal current density
(counting as two modes with $\omega = \pm cq$)
and three diffusive modes (with $\omega = iD q^2$) - energy and two
transverse current density modes.  To see this, consider the continuity equation,    
\begin{equation}
\frac{\partial \rho}{\partial t} + \vec{\nabla}.\vec{j} = 0 ,
\label{dense}
\end{equation}
and the Navier Stokes equation for current density conservation
linearized for small current density:
\begin{equation}
\frac{\partial \vec{j}}{\partial t} + \frac{1}{m}\vec{\nabla}p = \nu \nabla^2 \vec{j} ,
\label{mtm}
\end{equation}
where $p$ is the pressure , $m$ the mass of the constituent He-4 particle, 
and $\nu$ the kinematic viscosity.
One can also define the fluid velocity field $\vec{v}$,
which satisfies $\vec{j}= \rho \vec{v}$.
Upon taking the divergence of Eqn.\ref{mtm}, combining with the continuity
equation and linearizing for small current density and density variations,
one obtains first sound with velocity, $c = \sqrt{\frac{1}{m}\frac{\partial p}{\partial \rho}|_s}$.  Eqn.\ref{mtm} also implies two diffusive transverse
current density modes.
In addition, the energy density $\epsilon$ is
conserved:
\begin{equation}
\frac{\partial \epsilon}{\partial t} + \vec{\nabla}.\vec{Q} = 0,
\label{encon}
\end{equation}
where $\vec{Q}$ is the heat current. Upon 
 combining Eqn.\ref{encon} with the expression that defines
the thermal conductivity,
one obtains the heat diffusion equation,
\begin{equation}
\partial_t \epsilon = D_T \nabla^2 \epsilon \end{equation}
where we have used $C_V= \partial \epsilon / \partial T$ and the Einstein relation relating $\kappa$ and the thermal diffusion coefficient, $D_T$.

On approaching the $\lambda$-transition
into the superfluid, one must augment this hydrodynamics
with the slow dynamical relaxation of the order parameter.
This leads to six hydrodynamic modes in the superfluid
phase:  first sound, second sound (involving the order
parameter and predominantly energy) and the two diffusive
transverse current density modes.  In the superfluid
one adopts a "two-fluid" description,
in which the total density is decomposed into a superfluid and a normal
fluid component:
$\rho=\rho_s + \rho_n$.
Moreover, one
introduces a second velocity field - the superfluid velocity
$\vec{v}_s$, in addition to the "normal" fluid velocity
denoted $\vec{v}_n$, and the total
 current density is decomposed as
$\vec{j} = \rho_s \vec{v}_s + \rho_n \vec{v}_n $. 
The superfluid velocity field is taken to satisfy  
a Josephson relation,  
\begin{equation}
m \frac{\partial \vec{v_s}}{\partial t} + \vec{\nabla}\mu = 0,
\label{sfvelo}
\end{equation}
where $\mu$ is the chemical potential\cite{LL}.
Diffusion of energy in the normal fluid is replaced by second sound
oscillations of normal and superfluid. As the superfluid carries no
entropy, the oscillations involve an entropy wave associated with the 
normal fluid. This is easiest seen in the absence of dissipation
($\nu = 0$, $\kappa = 0$),
where conservation of entropy density $s$, maybe expressed in terms of the
associated entropy current $s \vec{v}_n$ carried by the normal fluid as
\begin{equation}
\frac{\partial s}{\partial t} + \vec{\nabla}.(s \vec{v}_n) = 0.
\label{entropy}
\end{equation}
One can  combine Eq.\ref{dense},\ref{mtm},\ref{sfvelo},\ref{entropy},
 and use the
fact that local changes in chemical potential are related to local thermal
gradients via the thermodynamic relation
\begin{equation}
\rho d \mu = -s d T + d p .
\label{thermodyn}
\end{equation} 
One then finds the superfluid velocity and the entropy 
participating in second sound motion with velocity
$c_s = \sqrt{\frac{1}{\rho m}\frac{\rho_s s^2}{\rho_n (\frac{\partial s}{\partial T})}}$.

First sound and the two diffusive transverse current density modes
are also present in the superfluid phase, giving the anticipated
six hydrodynamic modes.

Turning next to the critical dynamics at the $\lambda$-transition,
one needs to take into account purely those hydrodynamic modes 
which go soft.  Coming from the superfluid side, the only such mode is
the second-sound mode since its velocity vanishes
along with the superfluid density.  The first sound velocity remains finite
through the transition, so that first sound oscillates much more
quickly than second sound at the same wavevector.  One thereby argues
that first sound should not enter into the low frequency critical
dynamics. Moreover, couplings of transverse current density modes are found to be 
irrelevant\cite{HH}. Thus for the critical dynamics, one
can focus on the conserved heat density and the order parameter,
which together comprise second sound in the superfluid.
More precisely, the non-dissipative coupling of energy and the 
order parameter below $T_c$, which results in second sound, is 
captured by the Poisson relationship
\begin{equation}
\{ \psi , M \} \sim \psi,
\label{conjug}
\end{equation}
where $M$ is the net combination of the energy and particle number 
densities taking part in the 
oscillation, and $\psi$ is the complex order parameter. The critical dynamics
of such a system is described by a generalized
time-dependent Ginzburg-Landau type model involving
these two fields, and was denoted as Model F in Ref.\cite{HH}.

\section{Coulomb interactions and impurities}

Turning to the case of the charged, pure fluid, which is appropriate 
in the normal phase for  a pure metal,
the presence of an electric
field $\vec{E}$ gives rise to acceleration. Thus, the r.h.s of Eq.\ref{mtm}
acquires a term $\frac{e}{m}\rho\vec{E}$, and that of Eq.\ref{sfvelo}
acquires a term $e \vec{E}$, where '$e$' is the charge of the 
electron. In addition, distortions of the charged fluid itself give
rise to an electric field described by Poisson's equation,
\begin{equation}
\vec{\nabla}.\vec{E} = 4 \pi e \delta \rho ,
\label{fish}
\end{equation}
where $\delta \rho$ corresponds to density distortions.
Consequently, in three dimensions, the sound mode involving
density and longitudinal current density
becomes a high-energy plasmon with frequency $\omega_p = 
\sqrt{\frac{4 \pi e^2 \rho}{m}}$ in the limit $\vec{q} \to 0$,
and thus drops out of the hydrodynamic description.
The transverse diffusive current density modes remain.
 Importantly, even with the
modifications associated with the presence of charge, one  finds the
diffusive heat mode above $T_c$.

Below $T_c$, in the pure three dimensional superconductor,
the low frequency, long wavelength 
second sound mode survives because it involves no net density distortions.
To see this, we note that due to the gapped plasmon the density
is not a hydrodynamic variable, so one can effectively set $\delta \rho =0$.
Upon linearizing the continuity equation Eq.\ref{dense} for small velocities,
this implies that $\nabla \cdot \vec{v}_n = -(\rho_s /\rho_n) \nabla \cdot \vec{v}_s$.  Upon inserting this into the linearized entropy
conservation Eqn.\ref{entropy}, and combining with the divergence
of the Josephson relation Eq.\ref{sfvelo}, one finds that
\begin{equation}
\frac{\partial^2 s}{\partial t^2} + \frac{\rho_s s}{\rho_n m}  \nabla^2 \mu = 0.
\end{equation}
Finally, using the thermodynamic relation Eqn.\ref{thermodyn}
one arrives  at the wave equation for the entropy density
propogating once more with velocity $c_s = \sqrt{\frac{1}{\rho m}\frac{\rho_s s^2}{\rho_n (\frac{\partial s}{\partial T})}}$.
In addition to second sound, one expects the two diffusive
 transverse current density modes
to be present in the charged superfluid, just as in He-4. 

Thus, essentially the sole effect of Coulomb interactions on the superfluid
hydrodynamics is the conversion of first sound into the non-hydrodynamic
plasmon mode.  Since first sound was argued in any event to decouple
from the critical dynamics at the $\lambda$-transition, the critical
dynamics in the charged superfluid is expected to be described by the
same theory - that is model F, except with the density $M$
referring to pure energy.

Consider next
the case of an uncharged fluid in the presence of impurities
(eg. He-4 absorbed in a porous medium).
Impurities violate momentum conservation
and this leads to a dramatic modification of the
hydrodynamics.  In particular, with only density and
energy being conserved, one expects two hydrodynamic modes
above $T_c$ and three below.
Absence of momentum
conservation can be explicitly captured by
\begin{equation}
\frac{\partial \vec{j_n}}{\partial t} = - \frac{\vec{j_n}}{\tau},
\label{mtmdecay}
\end{equation}
where $1/\tau$ is the decay rate of  current density $\vec{j_n}$.
Then, above $T_c$, the first sound mode
involving longitudinal current density and density cannot propagate
at low frequencies. It is replaced by a damped current density mode
($\omega = -i/\tau$) and a diffusive
density mode with diffusion constant $D_{\rho} = c^2 \tau$, where '$c$' is the velocity
of first sound in the pure case.  Energy continues to be a diffusive
mode in the presence of impurities.

Below $T_c$, the uncharged impure system once again has a superfluid component
moving with velocity  $v_s$. The superfluid can couple to density in a
non-dissipative fashion,
but not to entropy, which is not even conserved, and is
still mainly associated with  the normal
fluid that can no longer propagate ballistically.  
We can thus simply drop the normal fluid velocity from the continuity equation
Eqn.\ref{dense},
which upon linearization and combination with the Josephson relation gives,
$\partial_t^2 \rho = (\rho_s/m) \nabla^2 \mu$.  This describes
a fourth sound mode involving purely superfluid oscillations
propagating with velocity
$c_{imp} = \sqrt{\frac{\rho_s}{m (\frac{\partial \rho}{\partial \mu})}}$.
In addition, 
the diffusive energy mode persists below
$T_c$.

To model the critical dynamics one needs to re-interpret
the conserved density $M$ in model F as the particle number density, $\rho$,
and then augment the model with an additional conserved
energy density which is diffusive both above and below $T_c$.
The conserved energy density will be coupled to the order
parameter $\psi$ via a term in the free energy of the
form $\epsilon |\psi|^2$.  This will lead to a coupling
in the resulting time-dependent equations of motion.

Finally, we arrive at the case of interest -the impure superconductor-,
which we model by the charged, impure superfluid. The hydrodynamic modes
may be obtained by incorporating the modifications described
above for charge and impurities to the linearized hydrodynamic equations
of the He-4 system. The normal state describes an impure metal.
Due to the long-ranged Coulomb interactions
the density, although conserved, is gapped
up at the plasma frequency and thus again drops
out of the hydrodynamic description.
Moreover, with impurities,
the transverse current density modes are also
damped. So above $T_c$, in this system of the impure metal,
associated with the only remaining 
hydrodynamic variable one has a diffusive thermal mode. 

Below $T_c$,
superconducting order gives rise to a second mode. As in the case
of the uncharged, impure fluid, the order parameter cannot couple
non-dissipatively to heat. But unlike in this case, it cannot even do
so with density, which as in the pure charged case, is no longer
a hydrodynamic variable since its distortions cost
electrostatic energy. As a result, 
there exist two {\it diffusive} modes related to the energy and order parameter,
and no oscillatory hydrodynamic modes\footnote{Oscillatory modes involving
normal and superfluid densities were  observed in aluminium
films\cite{CG}. However, modified hydrodynamic treatments\cite{Bray}
show that except for a narrow window
of momenta, these Carlson-Goldman modes are damped.}.

While our hydrodynamic treatment employed simple linearized equations
appropriate for mode analysis, an exhaustive study along the lines of
the case of He-4\cite{Khalat} may prove useful for the case of the impure
superconductor.

\section{Criticality}

We are now equipped to model the critical dynamics of the dirty superconductor.
We have seen that above $T_c$, there only exists a diffusive energy mode.
Below $T_c$, in striking contrast to superfluid $He^4$, instead of a 
second sound oscillatory mode, there exists one diffusive mode associated with
conserved energy density, and one diffusive mode associated with the order
parameter, which takes the form of a complex scalar.
The simplest phenomenological model incorporating these ingredients,
and all possible relevant couplings, is described by Model C of Ref.\cite{HH}
, and is analyzed in detail for its critical dynamics in Ref.\cite{HHM}.
It is defined by the following set of equations of motion involving
the complex order parameter $\psi$ and the conserved energy density:
\begin{eqnarray}
\frac{\partial \psi}{\partial t} & = &-\Gamma_0 \frac{\delta F_0}{\delta \psi^*}  + \eta, \\
\frac{\partial \epsilon}{\partial t} & = &-\kappa_0 \nabla^{2} (\frac{\delta F_0}{\delta \epsilon} - \delta) + \zeta, \\
F_0 & = & \int d^dx(\frac{1}{2} r_0 |\psi|^2 + u_0 |\psi|^4 + \frac{1}{2}|\vec{\nabla} \psi|^2 + \gamma_0 |\psi|^2 \epsilon \nonumber \\
&  + & \frac{1}{2} C_0^{-1}\epsilon^2).
\label{eom}
\end{eqnarray}
Here $\Gamma_0$ and $\kappa_0$ are the
bare transport co-efficients for the order parameter and the energy, respectively.  An external source field, $\delta$ has been included
in Eqn.\ref{eom},
and 
$\eta$ and $\zeta$ are Langevin noise sources appropriate 
for $\psi$  (complex) and
$\epsilon$ (real), respectively. The energy density and the superconducting order parameter are coupled together
via the coupling constant $\gamma_0$.

In equilibrium, $\psi$ and $\epsilon$
minimize the functional $F_0$.
In a functional integral formulation applicable for the equilibrium
distribution\cite{HHM}, one can integrate over the energy field, $\epsilon$. The
resulting functional of the order parameter $\psi$ displays the statics
of 3D XY critical behaviour, appropriate
for an extreme type II superconductor(see for e.g. Ref\cite{FFH,critex}).
The Harris criterion (see for e.g. Ref\cite{Goldenfeld}) shows that
this holds true even in the presence of disorder.
Thus, the pure 3D XY model provides an appropriate description
for critical static properties of the dirty superconductor.

We next turn to the critical dynamics.  To obtain the thermal conductivity, one first defines
the energy density linear response function
 $\chi(\vec{q},\omega)$, by the relation
\begin{equation}
<\epsilon(\vec{q},\omega)>_{\delta} = \chi(\vec{q},\omega) \delta(\vec{q},\omega).
\label{response}
\end{equation}
Quite generally, in such a diffusive system, one expects this response function at low frequencies and wavevectors to
take the standard form:
\begin{equation}
\chi(\vec{q},\omega) = \frac{\kappa q^2}{-i \omega + D_T q^2},
\label{fullesp}
\end{equation}
where $\kappa$ is the thermal conductivity and $D_T$ the thermal diffusion constant.  Thus, the thermal conductivity can be extracted as,
\begin{equation}
\kappa = \lim_{\vec{q} \to 0}q^{-2}\Big[\frac{i \partial \chi^{-1}(\vec{q},\omega)}{\partial \omega}\Big|_{\omega=0}\Big]^{-1}.
\label{thermcond}
\end{equation}

In the absence of coupling to the order parameter, $\chi$ can be readily
computed from the (linear) equation of motion for $\epsilon$,
which gives,
\begin{equation}
\chi^{-1}_0(\vec{q},\omega) = \frac{-i \omega}{\kappa_0 q^2} + C_0^{-1} ,
\label{baresp}
\end{equation}
which has the general diffusive form as in Eqn.\ref{fullesp}
with the identifications $\kappa = \kappa_0$ and $D_T = \kappa_0/C_0$.

The effect of the fluctuating
order parameter on the thermal conductivity $\kappa$, can be studied
by treating the couplings $u_0$ and $\gamma_0$ perturbatively. 
Specifically, one can set up a perturbative expansion
for the "self-energy" ($\Sigma$) which modifies the thermal response function:
\begin{equation}
\chi^{-1}(\vec{q},\omega) = \chi^{-1}_0(\vec{q},\omega) + \Sigma(\vec{q},\omega) .
\end{equation}
As is clear from Eqn.\ref{thermcond}, a renormalization of the bare thermal
conductivity $\kappa_0$ requires a contribution
to the self energy of the form $\Sigma \sim -i\omega/q^2$,
which is divergent as $q \rightarrow 0$ for fixed non-zero $\omega$.
But as argued by Ref.\cite{HH,HHM}, away from criticality,
with $r_0 > 0$, the self energy will be finite
in the $q \rightarrow 0$ limit at all orders in perturbation theory,
being protected in the infrared by $r_0$ and in the ultraviolet
by a high momentum cutoff.  Thus, these perturbations
can generate 
no corrections
to $\kappa$, and the thermal conductivity will
be given exactly by the "bare" parameter $\kappa_0$\cite{HHM}.
Since $\kappa_0$ is a coupling constant which depends
on short distance physics only, it will necessarily be
a smooth
function of temperature.  This thereby establishes
that the thermal conductivity should be non-singular
 and smooth as one cools through the 
superconducting transition.

It is worth emphasizing that although the thermal conductivity
 itself is non-singular at the transition, critical singularities 
driven by the order parameter flucutations will generally 
enter into
the dynamical relaxation of the energy. 
Smoothness of the thermal conductivity is
intimately tied to the fact that it is a zero-frequency,
zero-wavevector quantity.  Indeed, the dependence
of $\kappa$ on wavevector or frequency is expected to be singular
at the critical point.

To recapitulate and emphasize, fluctuations in the superconducting order 
parameter will certainly affect the time dependent
relaxation of the energy. However, the thermal conductivity - the transport 
co-efficient associated with energy propagation - remains smooth and 
finite across the transition.

Warm thanks to L. Balents, E. Boaknin, N.P. Ong, L. Taillefer and 
A.A. Varlamov for their comments and correspondences.
This research was supported by NSF Grants DMR-97-04005,
DMR95-28578
and PHY94-07194.

\end{multicols}

\end{document}